\documentclass[10pt]{article}
\usepackage{slashed}
\usepackage{amsmath}
\usepackage{amssymb}
\usepackage{axodraw}
\newcommand{\be}{\begin{equation}}
\newcommand{\ee}{\end{equation}}
\newcommand{\ba}{\begin{eqnarray}}
\newcommand{\ea}{\end{eqnarray}}
\newcommand{\rmi}[1]{{\mbox{\scriptsize #1}}}

\newcommand{\tr}{{\rm Tr\,}}
\newcommand{\nn}{\nonumber \\}
\newcommand{\fr}[2]{{\frac{#1}{#2}\,}}
\newcommand{\msbar}{{\overline{\mbox{\rm MS}}}}

\renewcommand{\(}{\left(}
\renewcommand{\)}{\right)}

\newcommand{\e}{\epsilon}

\newcommand{\mLARGE}[1]{\hbox{\LARGE $#1$}}
\def\sumint{\hbox{$\sum$}\!\!\!\!\!\!\!\int}
\newcommand{\I}{{\cal I}^0}
\newcommand{\It}{\widetilde{\cal I}^0}
\newcommand{\Taut}{\widetilde{\mLARGE{\tau}}}
\newcommand{\Mt}{\widetilde{\cal M}}
\newcommand{\N}{{\cal N}}
\renewcommand{\ln}{{\rm ln}}
\newcommand{\mubar}{\bar{\mu}}

\newcommand{\aM}[1]{\alpha_\rmi{M#1}}
\newcommand{\aG}{\alpha_\rmi{G}}

\newcommand{\aE}[1]{\alpha_\rmi{E#1}}

\newcommand{\imathb}{i}

\makeatletter
\renewcommand\section{\@startsection {section}{1}{\z@}%
                                   {-5.5ex \@plus -1ex \@minus -.2ex}
                                   {2.3ex \@plus.2ex}%
                                   {\normalfont\large\bfseries}}
\renewcommand\subsection{\@startsection{subsection}{2}{\z@}%
                                     {-3.25ex\@plus -1ex \@minus -.2ex}%
                                     {1.5ex \@plus .2ex}%
                                     {\normalfont\normalsize\bfseries}}
\renewcommand\thesection {\@arabic\c@section}
\renewcommand\thesubsection   {\thesection.\@arabic\c@subsection}
\renewcommand{\@seccntformat}[1]{%
\csname the#1\endcsname.\hspace{1.0em}}
\makeatother

\newcommand{\pic}[1]{\;\parbox[c]{30pt}{\begin{picture}(30,30)(0,0)
\SetWidth{1.0}\SetScale{1.0} #1 \end{picture}}\;}

\newcommand{\picb}[1]{\;\parbox[c]{48pt}{\begin{picture}(45,30)(-9,0)
\SetWidth{1.0}\SetScale{1.0} #1 \end{picture}}\;}
\newcommand{\picc}[1]{\;\parbox[c]{45pt}{\begin{picture}(45,30)(0,0)
\SetWidth{1.0}\SetScale{1.0} #1 \end{picture}}\;}

\def\Lwidth{1}

\def\Agl(#1,#2)(#3,#4,#5){\PhotonArc(#1,#2)(#3,#4,#5){\Lwidth}
{6.283 #3 mul 360 div #4 #5 sub #4 #5 sub mul sqrt mul Ldensity mul}}
\def\Lgl(#1,#2)(#3,#4){\Photon(#1,#2)(#3,#4){\Lwidth}
{#1 #3 sub #1 #3 sub mul #2 #4 sub #2 #4 sub mul add sqrt Ldensity mul}}
\def\Agh(#1,#2)(#3,#4,#5){\DashArrowArc(#1,#2)(#3,#4,#5){1}}
\def\Aagh(#1,#2)(#3,#4,#5){\DashArrowArcn(#1,#2)(#3,#5,#4){1}}
\def\Lgh(#1,#2)(#3,#4){\DashArrowLine(#1,#2)(#3,#4){1}}
\def\Lagh(#1,#2)(#3,#4){\DashArrowLine(#3,#4)(#1,#2){1}}
\def\Ahh(#1,#2)(#3,#4,#5){\DashCArc(#1,#2)(#3,#4,#5){1}}
\def\Lhh(#1,#2)(#3,#4){\DashLine(#1,#2)(#3,#4){1}}
\def\Aqu(#1,#2)(#3,#4,#5){\ArrowArc(#1,#2)(#3,#4,#5)}
\def\Aaqu(#1,#2)(#3,#4,#5){\ArrowArcn(#1,#2)(#3,#5,#4)}
\def\Lqu(#1,#2)(#3,#4){\ArrowLine(#1,#2)(#3,#4)}
\def\Laqu(#1,#2)(#3,#4){\ArrowLine(#3,#4)(#1,#2)}
\def\Aqq(#1,#2)(#3,#4,#5){\CArc(#1,#2)(#3,#4,#5)}
\def\Lqq(#1,#2)(#3,#4){\ArrowLine(#1,#2)(#3,#4)}
\def\Asc(#1,#2)(#3,#4,#5){\ArrowArc(#1,#2)(#3,#4,#5)}
\def\Lsc(#1,#2)(#3,#4){\ArrowLine(#1,#2)(#3,#4)}
\def\DAsc(#1,#2)(#3,#4,#5){\DashCArc(#1,#2)(#3,#4,#5){3}}
\def\DLsc(#1,#2)(#3,#4){\DashLine(#1,#2)(#3,#4){3}}
\def\TAsc(#1,#2)(#3,#4,#5){\SetWidth{2.0}\CArc(#1,#2)(#3,#4,#5)\SetWidth{1.0}}
\def\TLsc(#1,#2)(#3,#4){\SetWidth{2.0}\ArrowLine(#1,#2)(#3,#4)\SetWidth{1.0}}

\makeatletter \@addtoreset{equation}{section} \makeatother
\renewcommand{\theequation}{\arabic{section}.\arabic{equation}}
\makeatletter
\renewcommand\section{\@startsection {section}{1}{\z@}%
                                   {-5.5ex \@plus -1ex \@minus -.2ex}
                                   {2.3ex \@plus.2ex}%
                                   {\normalfont\large\bfseries}}
\renewcommand\subsection{\@startsection{subsection}{2}{\z@}%
                                     {-3.25ex\@plus -1ex \@minus -.2ex}%
                                     {1.5ex \@plus .2ex}%
                                     {\normalfont\normalsize\bfseries}}
\renewcommand\thesection {\@arabic\c@section}
\renewcommand\thesubsection   {\thesection.\@arabic\c@subsection}
\renewcommand{\@seccntformat}[1]{%
\csname the#1\endcsname.\hspace{1.0em}}
\makeatother

\begin{document}

\begin{titlepage}
\begin{flushright}
hep-ph/0212283\\
HIP-2002-66/TH\\
\end{flushright}
\begin{centering}
\vfill

{\large {\bf Quark number susceptibilities of hot QCD up to ${\mbox{\boldmath${\rm g}^6\ln\,{\rm g}$}}$}}

\vspace{0.8cm}

{A. Vuorinen\footnote{aleksi.vuorinen@helsinki.fi}}

\vspace{0.8cm}

{\em
Department of Physical Sciences,
Theoretical Physics Division \\
P.O. Box 64,
FIN-00014 University of Helsinki,
Finland\\}

\vspace*{1.4cm}

\end{centering}

\noindent

The pressure of hot QCD has recently been determined to the last perturbatively computable order $g^6\ln\,g$  by Kajantie {\em et al.} \cite{klry} using three-dimensional effective theories. A similar method is applied here to the pressure in the presence of small but non-vanishing quark chemical potentials, and the result is used to derive the quark number susceptibilities in the limit $\mu = 0$. The diagonal quark number susceptibility of QCD with $n_f$ flavours of massless quarks is evaluated to order $g^6\ln\,g$ and compared with recent lattice simulations. It is observed that the results qualitatively resemble the lattice ones, and that when combined with the fully perturbative but yet undetermined $g^6$ term they may well explain the behaviour of the lattice data for a wide range of temperatures.

\vfill
\noindent

\vspace*{1cm}

\noindent

\vfill

\end{titlepage}


\section{Introduction}
The grand potential is the most fundamental function describing the equilibrium properties of a thermodynamic system. By itself it amounts to minus the pressure $p(T,\mu_1,\mu_2,...)$ times the volume and from its derivatives one immediately obtains such quantities as entropy, specific heats, number densities and susceptibilities. In quantum chromodynamics a lot of attention has been devoted to determining the pressure in the quark-qluon plasma phase and e.g. lattice simulations have been successfully applied to the problem. They are, however, only available for temperatures no more than a few times above the critical temperature $T_c$ of the deconfinement phase transition and have only recently been extended to finite baryon densities.

So far the most powerful tool available for analytic calculations in QCD has been perturbation theory, the use of which is justified by the small value of the coupling constant $g$ at high energy densities due to asymptotic freedom. At $\mu=0$ the perturbative series of the pressure has recently been extended to order $g^6\ln\,g$ \cite{klry}, which marks the final step in a series of impressive computations starting from the determination of the order $g^2$ contribution \cite{es} and leading through the orders $g^3$ \cite{jk}, $g^4\ln\,g$ \cite{tt}, $g^4$ \cite{az} and $g^5$ \cite{zk,bn1}. The next $\mathcal{O}(g^6)$ term in the series is already out of reach for analytic computations due to infrared problems \cite{linde}.

At non-zero quark chemical potentials the pressure is at present known only to order $g^4\ln\,g$ \cite{tt} at finite $T$ and to order $g^4$ at $T=0$ \cite{fmcl} reflecting in part the computational complications induced by a finite value of $\mu$. From these results one may derive the quark number susceptibilities $\chi$ defined as second derivatives of the pressure with respect to the chemical potentials. They are both important and very interesting quantities, since they describe the effects of finite density being at the same time directly measurable on a lattice in the limit $\mu \rightarrow 0$ (see e.g. \cite{gup2,gup3,bern}). The perturbative results for $\chi$ obtained from \cite{tt} in this limit are, however, unable to produce even the qualitative behaviour of the lattice data. Recent HTL computations (see e.g. \cite{bir}) have improved the situation somewhat.

In the present paper the pressure of quark-gluon plasma with massless quarks is computed to order $g^6\ln\,g$ in the limit of small but non-vanishing chemical potentials $0\leq \mu/T \ll 1$. The calculation is a generalization of the recent $\mu=0$ paper \cite{klry} and applies the same method, matching of effective three-dimensional theories to four-dimensional QCD \cite{dr}, in separating the contributions of different momentum scales to $p_\rmi{QCD}$. The result is used in deriving the quark number susceptibilities at $\mu=0$ to order $g^6\ln\,g$, and the diagonal susceptibility is subsequently analyzed and compared with results of lattice simulations. One observes that the results of the present computation follow the same trend as the lattice ones, but that the effects of the yet undetermined contributions may still affect their behaviour considerably.

The non-trivial part of deriving the susceptibilities is the evaluation of the fermionic three-loop diagrams of full QCD that contribute to the pressure. To do that one applies here the results of yet unpublished work \cite{av}, in which the necessary diagrams are computed for arbitrary $T$ and $\mu$. In this paper the results are, however, only quoted as expanded to second order with respect to $\mu/T$, which is sufficient for the
present purposes. Due to the use of the effective theory approach one may here perform the calculations without any resummations applying dimensional regularization in the $\msbar$ scheme is to regulate both ultraviolet and infrared divergences. All fields are considered massless, while the chemical potentials of the quark flavours are regarded as being independent and non-zero and the temperature higher than $T_c$.

The general setup of the paper is presented in section 2, while the results for the pressure are assembled in section 3. Section 4 is then devoted to explaining the computation of the matching coefficients needed in deriving the pressure, and the diagonal quark number susceptibility at $\mu=0$ is  discussed in section 5. Conclusions are finally drawn in section 6 and the values of the matching coefficients listed in Appendix A.


\section{Setup and notation}
In Euclidean metric quantum chromodynamics is defined by the Lagrangian density
\ba
{\cal L}_\rmi{QCD} & = &  \fr14 F_{\mu\nu}^a F_{\mu\nu}^a + \bar\psi\slashed{D}\psi,
\ea
where
\ba
F_{\mu\nu}^a &=& \partial_\mu A_\nu^a - \partial_\nu A_\mu^a + g f^{abc} A_\mu^b A_\nu^c,\\
D_\mu &=& \partial_\mu - i g A_\mu \;\;\,=\;\;\, \partial_\mu - i g A_\mu^a T^a,
\ea
and where the massless quark fields have been combined into a multi-component spinor $\psi$. The $N_c\times N_c$ -matrices $T^a$ are here the generators of the fundamental representation of $SU(N_c)$, and the relevant group theory factors read
\ba
C_A \delta_{cd} & \equiv & f^{abc}f^{abd} \;\,\;=\;\,\; N_c \delta_{cd}, \\
C_F \delta_{ij} & \equiv & (T^a T^a)_{ij} \;\,\;=\;\,\; \fr{N_c^2-1}{2N_c}\delta_{ij}, \\
T_F \delta_{ab} & \equiv & \tr T^a T^b \;\,\;=\;\,\; \fr{n_f}{2} \delta_{ab}
\ea
and
\ba
D \delta_{cd} \;\;\,\equiv\;\;\, d^{abc}d^{abd}
\;\;\,=\,\;\;
 \fr{N_c^2-4}{N_c}\delta_{cd}.
\ea
The dimensions of the representations are $d_A \equiv \delta_{aa} =  N_c^2 - 1$ for the adjoint one and $d_F \equiv \delta_{ii} = d_A T_F/C_F = N_cn_f$ for the fermionic one.

The partition function is defined as a path integral over all fields of the functional
\ba
{\rm
exp}\bigg\{\!-\Big(S_\rmi{QCD}-\sum_f\mu_f N_f\Big)\bigg\} \;\;\,=\;\;\, {\rm
exp}\bigg\{\!-\!\int_0^{\beta} \! {\rm d}\tau \!\int \! {\rm d}^d x\, \big({\cal L}_\rmi{QCD}
-\psi^{\dagger}{\mbox{\boldmath$\mu$}}\psi\big)\bigg\},
\label{z}
\ea
where $\beta = 1/T$ and ${\mbox{\boldmath$\mu$}}$ is a diagonal $n_f\times n_f$ -matrix in flavour space representing the different chemical potentials of the quark flavours. In (\ref{z}) the space dimensionality is $d=3-2\e$ signifying the use of dimensional regularization. The partition function gives the pressure through the relation $p = T/V\,\ln\,Z$, where the infinite-volume limit is assumed.

At high temperatures and small chemical potentials $\mu \ll T$ the pressure of QCD can be separated into three parts $p_\rmi{QCD} =
p_\rmi{E}+p_\rmi{M}+p_\rmi{G}$ corresponding respectively to the different momentum scales $2\pi T$, $gT$ and $g^2T$
contributing to it \cite{dr,bn1}. By definition
\ba
p_\rmi{E}\!\(T,\mu\) \;\;\,\equiv\,\;\;
 p_\rmi{QCD}\!\(T,\mu\)-\fr{T}{V} \ln  \int {\cal D}A_i^a \,
{\cal D}A_0^a \exp\Big\{\!-\!S_\rmi{E}\Big\},
\ea
where $S_\rmi{E}$ is the action of a three-dimensional effective theory with the Lagrangian density \cite{hlp}
\ba
{\cal L}_\rmi{E} \;\;\, = \,\;\; \fr12 \tr F_{ij}^2 + \tr [D_i,A_0]^2 + m_\rmi{E}^2\tr A_0^2 + \fr{\imathb g^3}{3\pi^2}
\sum_f\mu_f \,\tr A_0^3 + \delta{\cal L}_\rmi{E},
\label{leqcd}
\ea
obtained from QCD by dimensional reduction \cite{klrs2}. In (\ref{leqcd}) $m_\rmi{E}$ and the coupling constant $g_\rmi{E}$ appearing
in $F_{ij}$ are parameters to be determined in full QCD, and terms that contribute to the pressure starting at order $g^6$ or higher have been assembled in $\delta{\cal L}_\rmi{E}$.

The functions $p_\rmi{M}\!\(T,\mu\)$ and $p_\rmi{G}\!\(T\)$ are similarly defined by
\ba
p_\rmi{M}\!\(T,\mu\) &\equiv& p_\rmi{QCD}\!\(T,\mu\)-p_\rmi{E}\!\(T,\mu\)-\fr{T}{V} \ln  \!\int {\cal D}A_i^a
\exp\Big\{\!-\!S_\rmi{M}\Big\} \nn
&\equiv& p_\rmi{QCD}\!\(T,\mu\)-p_\rmi{E}\!\(T,\mu\) -p_\rmi{G}\!\(T\), \\
{\cal L}_\rmi{M} & = & \fr12 \tr F_{ij}^2 + \delta{\cal L}_\rmi{M}
\ea
with yet another coupling constant $g_\rmi{M}$ appearing in the definition of $F_{ij}$. At leading order the different parts contribute to the pressure as $p_\rmi{E}\sim g^0$, $p_\rmi{M}\sim g^3$ and $p_\rmi{G}\sim g^6\ln\,g$. As is indicated above by writing $p_\rmi{G}\!\(T\)$, it will be seen that the dependence of $p_\rmi{G}$ on $\mu$ is of higher order than what is considered here.

The contributions of the different momentum scales to the $\mu=0$ pressure have been analyzed in detail in \cite{klry}. Since the effects of finite chemical potentials with few exceptions manifest themselves merely as changes in the matching coefficients defined there, the treatment of the pressure in the present paper will be restricted to quoting the results for the $p_\rmi{N}$'s from \cite{klry} and discussing the effects of finite $\mu$ on the coefficients.

The momentum integration measure and the shorthands for sum-integrals used from here onwards are
\ba
\int_p &\equiv& \int\! \fr{{\rm d}^{d} p}{(2\pi)^{d}} \;\;\,=\;\;\, \Lambda^{-2\e}\(\!\fr{e^{\gamma}\bar{\Lambda}^2}{4\pi}\!\!\)^{\!\!\e}\!\!\int\!
\fr{{\rm d}^{d} p}{(2\pi)^{d}}, \\
\sumint_{P/\{P\}} &\equiv& T \sum_{p_0/\{p_0\}} \int_p,
\ea
where $\bar{\Lambda}$ is the $\msbar$ scale and $p_0 \equiv 2n\pi T$ stands for bosonic and $\{p_0\} \equiv (2n+1)\pi T-i\mu$ for fermionic Matsubara frequencies. In the following sections the chemical potentials usually appear in the dimensionless combination
\ba
\mubar \equiv \fr{\mu}{2\pi T}.
\ea

\section{The pressure at $0\leq\mu \ll T$}
Collecting results from \cite{klry} and \cite{hlp} one may, in analogy with \cite{klry}, write the different parameters required in computing $p_\rmi{QCD}$ to order $g^6\ln\,g$ in the form
\ba
 \frac{p_\rmi{G}(T)}{T \Lambda^{-2 \epsilon} } &=&
 d_A C_A^3 \frac{g_\rmi{M}^6}{(4\pi)^4}\, \ln\frac{\bar{\Lambda}}{2m_\rmi{M}}
 \Big[8\, \aG + {\mathcal O}(\epsilon) \Big], \label{pg} \\
 \frac{p_\rmi{M}(T,\mu)}{T \Lambda^{-2 \epsilon} } & = &
 \frac{1}{(4\pi)}
 d_A  m_\rmi{E}^3
 \bigg[\fr13 + {\cal O}(\epsilon) \bigg] \nn
 & + &
 \frac{1 
 }{(4\pi)^2}
 d_A C_A
 g_\rmi{E}^2 m_\rmi{E}^2
 \bigg[-\frac{1}{4\epsilon} - \fr34 -\ln\frac{\bar{\Lambda}}{2 m_\rmi{E}}
 + {\cal O}(\epsilon) \bigg] \nn
 & + &
 \frac{1 
 }{(4\pi)^3}
 d_A C_A^2
 g_\rmi{E}^4 m_\rmi{E}
 \bigg[-\frac{89}{24} - \fr16 \pi^2 + \frac{11}{6}\ln\,2
 + {\cal O}(\epsilon) \bigg] \nn
 & + &
 \frac{1 
 }{(4\pi)^4}
 d_A C_A^3
 g_\rmi{E}^6\, \ln\frac{\bar{\Lambda}}{2 m_\rmi{E}}
 \Big[ 8\,\aM{1}
  + {\cal O}(\epsilon)
 \Big]    \nn
& + &
\frac{1 
}{(4\pi)^4}
d_A D T_F^2
g_\rmi{E}^6\,\ln\frac{\bar{\Lambda}}{2 m_\rmi{E}}
\Big[ 8\,\aM{2}
+ {\cal O}(\epsilon) \Big], \label{pm}   \\
m_\rmi{M} &=& C_A g_\rmi{M}^2,    \\
g_\rmi{M}^2 &=& g_\rmi{E}^2\Big[1+\mathcal{O}(g_\rmi{E}^2/m_\rmi{E})\Big], \\
\label{pe}
 \fr{p_\rmi{E}(T,\mu)}{T\Lambda^{-2\e}} & = & T^3 \bigg[\aE{1}
 + g^2
 \Big(\aE{2} + {\cal O}(\epsilon)\Big)
 + \frac{g^4}{(4\pi)^2}
 \Big(\aE{3} + {\cal O}(\epsilon)\Big)
 \bigg], \\
 m_\rmi{E}^2 & = & T^2 \bigg[ g^2
 \Big( \aE{4} +
 \aE{5} \epsilon + {\cal O}(\epsilon^2) \Big)
 + \frac{g^4}{(4\pi)^2}
 \Big( \aE{6} +
 {\cal O}(\epsilon) \Big)  \bigg], \hspace*{0.5cm}  \\
 g_\rmi{E}^2 & = & T \bigg[ g^2 + \frac{g^4}{(4\pi)^2}
 \Big( \aE{7} + {\cal O}(\epsilon) \Big) \bigg],
\ea
where the effects of finite $\mu$ can be explicitly seen only in the appearance of an additional term proportional to $\aM{2}$ in (\ref{pm}). Here $g^2$ is the renormalized coupling of full QCD, and the values of the matching coefficients $\alpha$ can be immediately obtained from the results of \cite{klry}, \cite{es}, \cite{bir} and \cite{hlp} with the exception of $\aE{3}$ and $\aE{5}$. While the latter is found almost trivially through a one-loop computation, determining the first one requires the evaluation of all three-loop diagrams of full QCD that contribute to the pressure. This procedure is explained in section 4.

Adding together (\ref{pg}), (\ref{pm}) and (\ref{pe}) one has obtained a compact expression for the perturbative expansion of the pressure up to $\mathcal{O}(g^6\ln\,g)$
\ba
 \frac{p_\rmi{QCD}(T,\mu)}{T^4 \Lambda^{-2 \epsilon} } & = &
 \frac{p_\rmi{E}(T,\mu) + p_\rmi{M}(T,\mu) + p_\rmi{G}(T)}{T^4 \Lambda^{-2 \epsilon} } \nn
 & = &
 g^0 \bigg\{ \aE{1} \bigg\}
  +
 g^2 \bigg\{ \aE{2} \bigg\}
  +
 \frac{g^3}{(4\pi)}
 \bigg\{ \frac{d_A}{3} \aE{4}^{3/2} \bigg\} \nn
 & + &
 \frac{g^4}{(4\pi)^2} \bigg\{
 \aE{3} - d_A C_A
 \bigg[
 \aE{4} \bigg(
 \frac{1}{4\epsilon} + \fr34 + \ln\frac{\bar{\Lambda}}{2 g T \aE{4}^{1/2}}
 \bigg)
 + \fr14 \aE{5} \bigg] \bigg\} \nn
 & + &  \frac{g^5}{(4\pi)^3} \bigg\{ d_A \aE{4}^{1/2} \bigg[
 \fr12 \aE{6} - C_A^2
 \bigg(
 \frac{89}{24} + \frac{\pi^2}{6} - \frac{11}{6} \ln\,2
 \bigg)
 \bigg] \bigg\}  \nn
 & + &  \frac{g^6}{(4\pi)^4} \bigg\{d_AC_A\Big(\aE{6}+\aE{4}\aE{7}\Big)\ln\Big[g \aE{4}^{1/2}\Big] -
8\,d_A C_A^3\bigg(
 \aM{1}\,\ln\Big[g \aE{4}^{1/2}\Big]
 + 2\,\aG\, \ln\Big[gC_A^{1/2}\Big] \bigg) \nn
&-& 8\,d_A D T_F^2\aM{2} \,\ln\Big[g\aE{4}^{1/2}\Big] \bigg\}
\label{pres}
\ea
with the values of the coefficients $\alpha$ listed in Appendix A and the pole of $\aE{3}$ exactly cancelling the $1/\e$ term appearing in the order $g^4$ contribution. One should notice here that the numerical factors appearing inside the logarithms of the last $g^6\ln\,g$ term can be unambiguously defined only after the full order $g^6$ contribution has been determined.

\def\Elmeri(#1,#2,#3){{\pic{#1(15,15)(15,0,180)%
 #2(15,15)(15,180,360)%
 #3(0,15)(30,15)}}}

\def\Petteri(#1,#2,#3,#4,#5,#6){\pic{#3(15,15)(15,-30,90)%
 #1(15,15)(15,90,210)%
 #2(15,15)(15,210,330) #5(2,7.5)(15,15) #6(15,15)(15,30) #4(15,15)(28,7.5)}}

\def\Jalmari(#1,#2,#3,#4,#5,#6){\picc{#1(15,15)(15,90,270)%
 #2(30,15)(15,-90,90) #4(30,30)(15,30) #3(15,0)(30,0) #5(15,0)(15,30)%
 #6(30,30)(30,0) }}

\def\Oskari(#1,#2,#3,#4,#5,#6,#7,#8){\picc{#1(15,15)(15,90,270)%
 #2(30,15)(15,-90,90) #4(30,30)(15,30) #3(15,0)(30,0) #6(15,0)(15,15)%
 #5(15,15)(15,30) #8(30,30)(30,15) #7(30,15)(30,0) }}

\def\Sakari(#1,#2,#3){\picb{#1(15,15)(15,30,150)%
#1(15,15)(15,210,330) #2(0,15)(7.5,-90,90) #2(0,15)(7.5,90,270) %
#3(30,15)(7.5,-90,90) #3(30,15)(7.5,90,270) }}

\def\Maisteri(#1,#2){\picb{#1(15,15)(15,0,150)%
#1(15,15)(15,210,360) #2(0,15)(7.5,-90,90) #2(0,15)(7.5,90,270) #1(37.5,15)(7.5,0,360) }}

\section{The computation of $\aE{3}$}
In order to find the matching coefficient $\aE{3}$ one needs to compute the diagrammatic expansion of the QCD pressure to three-loop order at finite $T$ and $\mu$, but without any resummations. Since the corresponding expansion has already been derived for $\mu=0$ \cite{az}, one may restrict the treatment here to the $\mu$-specific part, i.e. to the diagrams containing fermionic lines. These diagrams are shown in Fig. 3 and contribute to the expansion as
\ba
p_\rmi{ferm} \;\;\,=\;\;\, p_\rmi{ferm}^{0} + \fr{1}{2} I_a + \fr{1}{3} I_b + \fr{1}{4} I_c + \fr{1}{2} I_d + \fr{1}{4} I_e
+ \fr{1}{2} I_f + \fr{1}{4} I_g + \fr{1}{4} I_h,
\label{pferm1}
\ea
where $p_\rmi{ferm}^{0}$ represents the pressure of non-interacting quarks (see e.g. \cite{kap}).

\begin{figure}[t]
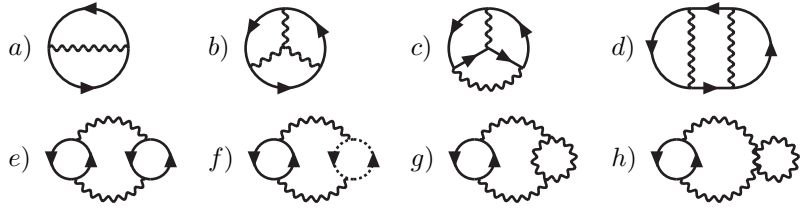

\centering
\ba \nonumber
\begin{array}{llll}
a)~ \Elmeri(\Asc,\Asc,\Lgl) & b)~
\Petteri(\Asc,\Asc,\Asc,\Lgl,\Lgl,\Lgl)
& c)~
\Petteri(\Asc,\Agl,\Asc,\Lsc,\Lsc,\Lgl)
& d)~
\Jalmari(\Asc,\Asc,\Lsc,\Lsc,\Lgl,\Lgl)
\nn
\nn
e)~
\Sakari(\Agl,\Asc,\Asc)
& f)~
\Sakari(\Agl,\Asc,\Agh)
& g)~
\Sakari(\Agl,\Asc,\Agl)
& h)~
\Maisteri(\Agl,\Asc)
\nn
\end{array}
\ea
\caption[a]{The two- and three-loop fermionic diagrams contributing to the pressure. The solid, wiggly and dashed lines stand respectively for the quark, gluon and ghost fields.}
\end{figure}

Applying the usual Euclidean space finite temperature Feynman rules to the individual diagrams, performing several shifts of integration momenta and taking advantage of the fact that the purely bosonic version of $\widetilde{\mLARGE{\tau}}$ vanishes at $\mathcal{O}(\e^0)$ \cite{az}, the diagrams may be rewritten in terms of the sum-integrals
\ba
\label{ints1}
{\cal I}_{n}^m &\equiv& \sumint_P \fr{\(p_0\)^m}{\(P^2\)^n}, \\
\widetilde{\cal I}_{n}^m &\equiv& \sumint_{\{P\}} \fr{\(p_0\)^m}{\(P^2\)^n}, \\
\widetilde{\mLARGE{\tau}} &\equiv& \sumint_{\{PQ\}}
\fr{1}{P^2 Q^2 \(P-Q\)^2},
\ea
\ba
\widetilde{\mLARGE{\tau}}' &\equiv& \sumint_{\{PQ\}}
\fr{p_0}{P^2 Q^2 \(P-Q\)^4}, \\
\widetilde{\cal M}_{m,n} &\equiv& \sumint_{\{PQR\}}
\fr{1}{P^2Q^2\(R^2\)^{m}\big(\(P-Q\)^2\big)^n\(P-R\)^2\(Q-R\)^2}, \\
{\cal N}_{m,n} &\equiv& \sumint_{\{PQ\}R}
\fr{1}{P^2Q^2\(R^2\)^{m}\big(\(P-Q\)^2\big)^n\(P-R\)^2\(Q-R\)^2},
\label{ints2}
\ea
defined here in analogy with \cite{bn1}. For example, for the diagram $b$ one obtains in the Feynman gauge
\ba
I_b &=& -\sumint_{\{PQR\}} {\rm Tr} \Big[\fr{1}{\slashed{P}}\(g\gamma ^{\mu}T_{ij}^{a}\)
\fr{1}{\slashed{Q}}\(g\gamma ^{\nu}T_{jk}^{b}\)\fr{1}{\slashed{R}}\(g\gamma ^{\rho}T_{ki}^{c}\)\Big] \imathb g f_{abc} \nn
&&\times \fr{g_{\mu\rho}\(2P_{\nu}-Q_{\nu}-R_{\nu}\)+g_{\rho\nu}\(2R_{\mu}-P_{\mu}-Q_{\mu}\)+
g_{\nu\mu}\(2Q_{\rho}-R_{\rho}-P_{\rho}\)}{\(P-Q\)^2\(Q-R\)^2\(R-P\)^2} \nn
&=&\fr{3}{2}d_A C_A T_F g^4 \sumint_{\{PQR\}} \fr{P_{\alpha}Q_{\beta}R_{\gamma}\(2P_{\nu}-Q_{\nu}-R_{\nu}\)}
{P^2Q^2R^2\(P-Q\)^2\(Q-R\)^2\(R-P\)^2}{\rm Tr}\Big[\gamma ^{\alpha}\gamma ^{\mu}\gamma ^{\beta}
\gamma ^{\nu}\gamma ^{\gamma}\gamma _{\mu}\Big] \nn
&=& 48 \(1-\e\)d_A C_A T_F g^4 \sumint_{\{PQR\}} \fr{P\cdot\(P-Q\) Q \cdot R}{P^2Q^2R^2\(P-Q\)^2\(Q-R\)^2\(R-P\)^2} \nn
&=& 12 \(1-\e\)d_A C_A T_F g^4 \Big[\({\cal I}_{1}^0-\widetilde{\cal I}_{1}^0\)\widetilde{\mLARGE{\tau}}+
\fr{1}{2} \widetilde{\cal M}_{0,0}\Big],
\label{diaga}
\ea
and the other ones produce in the same gauge
\ba
I_a &=& -4 \(1-\e\)d_A T_F g^2 \It_1\(\It_1-2\I_1\), \\
I_c &=& 4 \(1-\e\)d_A \(2C_F -C_A\) T_F g^4\Big[4\(\I_1-2\It_1\)\Taut + \(2+\e\)\N_{0,0}
 - 2\e\Mt_{0,0} + 2\N_{1,-1}\Big], \\
I_d &=& -8\(1-\e\)^2 d_A C_F T_F g^4 \Big[\(\I_1-\It_1\)^2\It_2-2\It_1\Taut+\Mt_{0,0}+\Mt_{1,-1}\Big], \\
I_e &=& -8d_A T_F^2 g^4 \Big[4\(1+\e\)(\It_1)^2\I_2 - 16\widetilde{\cal I}_{1}^1 \widetilde{\mLARGE{\tau}}'-
\(1-\e\)\(\N_{0,0}-4\It_1\Taut\) - 2\N_{1,-1}-2\N_{2,-2} \Big], \\
I_f &=& -d_A C_A T_F g^4 \Big[8\I_1\It_1\I_2 - 2\I_1\Taut + \Mt_{0,0} - 2\Mt_{-2,2}\Big], \\
I_g &=& 4 d_A C_A T_F g^4 \Big[4\(6-5\e\)\I_1\It_1\I_2
-\(7-6\e\)\I_1\Taut -\Big(\fr{3}{2}-2\e\Big)\Mt_{0,0}
 - \(5-4\e\)\Mt_{-2,2}\Big], \\
I_h &=& -8\(3-2\e\)\(1-\e\)d_A C_A T_F g^4 \Big[2\I_1\It_1\I_2-\I_1\Taut \Big].
\label{diagb}
\ea

Substituting Eqs. (\ref{diaga}) - (\ref{diagb}) into (\ref{pferm1}) and allowing explicitly flavour-dependent chemical potentials, one has obtained a representation for $p_\rmi{ferm}$ in terms of the sum-integrals defined above
\ba
\label{pferm2}
p_\rmi{ferm} &=& \fr{1}{n_f}\sum_f \bigg\{p_\rmi{ferm}^{0} - 2\(1-\e\)d_A T_F Z_{g}^{2} g^2 \It_1\(\It_1-2\I_1\) \nn
&+& d_A g^4\bigg( C_AT_F\(1-\e\)\Big[8\(1+\e\)\I_1\It_1\I_2-4\e\I_1\Taut+4\It_1\Taut \nn
&+& 2\e\Mt_{0,0}-\(2+\e\)\N_{0,0} - 2\N_{1,-1} - 4\Mt_{-2,2}\Big] \nn
&-& 2C_F T_F \(1-\e\)\Big[2\(1-\e\)\(\I_1-\It_1\)^2\It_2- 4\I_1\Taut+4\(1+\e\)\It_1\Taut \nn
&+& 2\Mt_{0,0}-\(2+\e\)\N_{0,0} + 2\(1-\e\)\Mt_{1,-1} - 2\N_{1,-1}\Big]\bigg)\bigg\} \nn
&-& \fr{1}{n_f^2}\sum_{f\,g}d_A g^4 T_F^2\bigg\{8\(1+\e\)\It_1[\mu_f]\,\It_1[\mu_g]\,\I_2
+ 4\(1-\e\)\(\It_{1}[\mu_f]\,\Taut[\mu_g]+\It_{1}[\mu_g]\,\Taut[\mu_f]\) \\
&-&16\(\widetilde{\cal I}_{1}^1[\mu_f]\, \widetilde{\mLARGE{\tau}}'[\mu_g] + \widetilde{\cal I}_{1}^1[\mu_g]\,
\widetilde{\mLARGE{\tau}}'[\mu_f]\) - 2\(1-\e\)\N_{0,0}[\mu_f,\mu_g] - 4\N_{1,-1}[\mu_f,\mu_g] -
4\N_{2,-2}[\mu_f,\mu_g] \bigg\}. \nonumber
\ea
The renormalization coefficient of the gauge coupling, $Z_g$, is
given by
\ba
Z_g^2 \;\;\,=\;\;\, 1-\fr{11C_A-4T_F}{3}\fr{g^2}{\(4\pi\)^2}\fr{1}{\e},
\ea
and in the latter sum of (\ref{pferm2}) the $\N_{n,-n}$'s depend on both $\mu_f$ and $\mu_g$ through the respective
fermionic momenta. The sum-integrals appearing in the result are computed\footnote{Recently there has appeared a new paper
\cite{antti}, in which one- and two-loop sum-integrals analogous to those considered in \cite{av} have been
independently calculated.} in \cite{av} for arbitrary $T$ and $\mu$, and thus the coefficient $\aE{3}$ is available by
simply combining (\ref{pferm2}) with the bosonic part of the strict perturbation expansion of the pressure (see e.g.
Eq. (31) of \cite{bn1}). The outcome of the computation expanded to second order in $\mubar$ is displayed in
Eq. (\ref{alphae3}) of Appendix A.

\section{Quark number susceptibilities at $\mu=0$}
From the expression (\ref{pres}) one may at once extract the quark number susceptibilities defined by
\ba
\chi_{ij} \;\;\,\equiv\,\;\; \fr{\partial^2 p}{\partial\mu_i \partial\mu_j} \;\;\,=\;\;\, \chi
\delta_{ij}+\widetilde{\chi}\(1-\delta_{ij}\),
\ea
where the symmetry between the massless quark flavours at $\mu = 0$ has been exploited in the latter equality in dividing the result explicitly into a diagonal and an off-diagonal part. The off-diagonal susceptibility $\widetilde{\chi}$ is already known to order $g^6\ln\,g$ \cite{bir}, but the result obtained here for the diagonal one $\chi$ is new. Specializing to the physical case $N_c=3$ one readily obtains
\ba
\label{qcdsusca}
\fr{\chi}{\chi_0}|_{\mu=0} &=& 1 - 2\,\fr{g^2\!\(\rho\)}{4\pi^2} +
8\sqrt{1+\fr{n_f}{6}}\(\!\fr{g^2\!\(\rho\)}{4\pi^2}\!\!\)^{\!\!3/2} \! +
12\(\!\fr{g^2\!\(\rho\)}{4\pi^2}\!\!\)^{\!\!2}\ln\,\fr{g^2\!\(\rho\)}{4\pi^2} \nn
&-& \fr{1}{36}\bigg\{12\(33-2n_f\)\ln\fr{e^{\gamma}\rho}{4\pi T} -
432\,\ln\Big[1+\fr{n_f}{6}\Big]+133+26\,n_f
+ 16\(17+2\,n_f\)\ln\,2\nn
&-& 432\gamma - 432\fr{\zeta'(-1)}{\zeta(-1)}\bigg\}\bigg(\!\fr{g^2\!\(\rho\)}{4\pi^2}\!\!\bigg)^{\!\!2} \nonumber
\ea
\begin{figure}[t]

\centerline{\epsfxsize=7.3cm \epsfbox{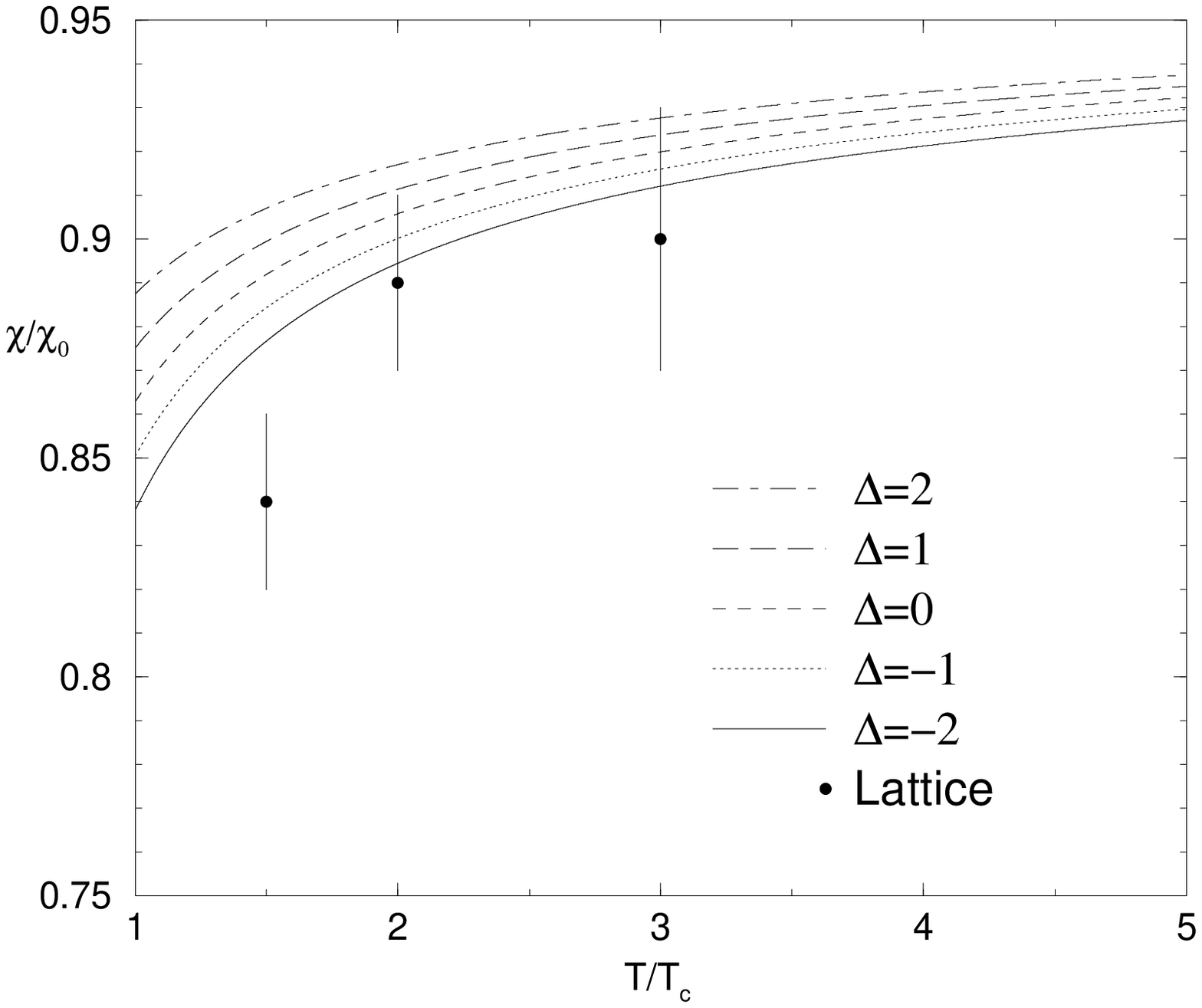}\;\;\;\;\;\;\;\epsfxsize=7.3cm \epsfbox{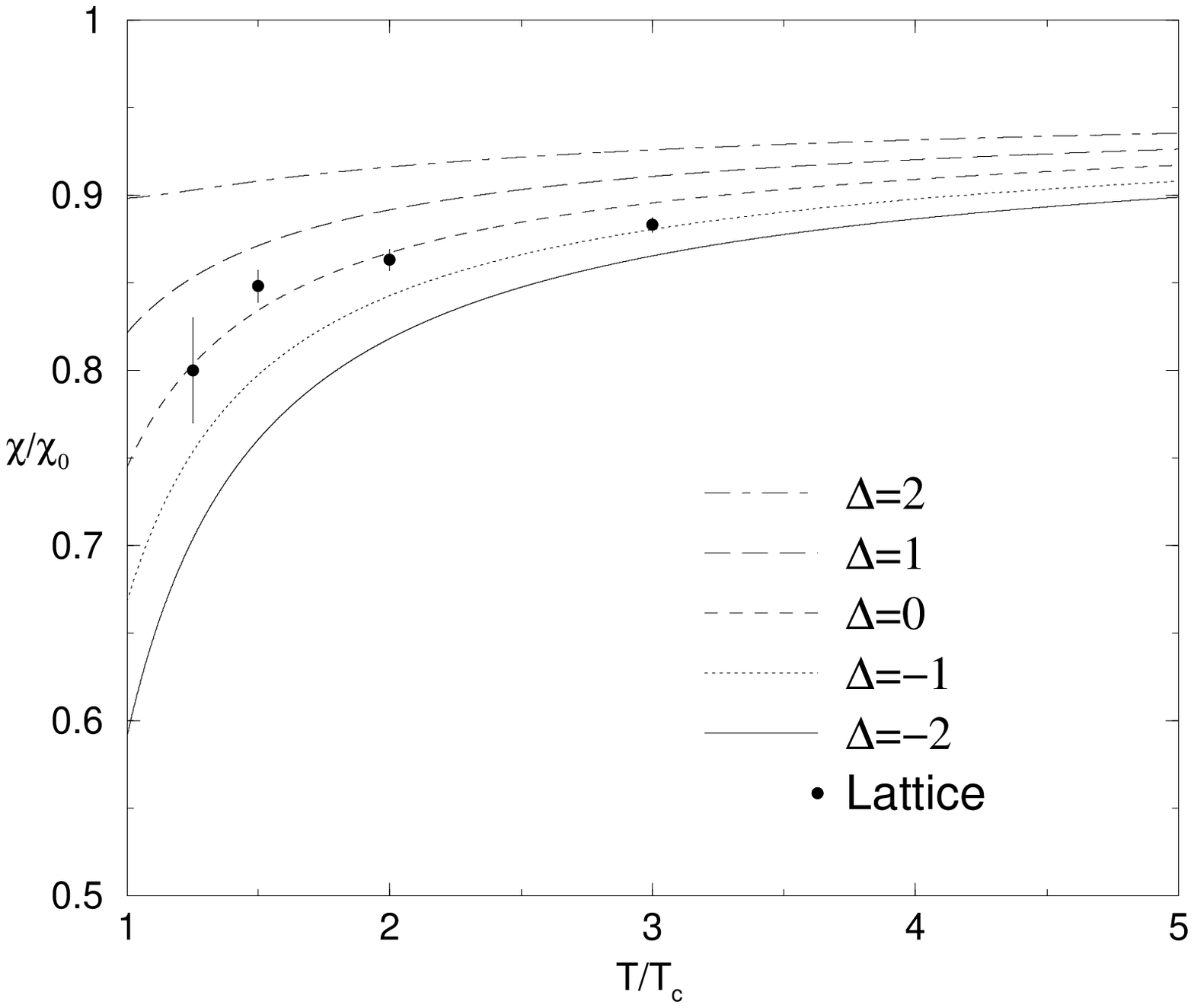}}

\caption[a]{$\chi/\chi_0$ plotted as a function of $T/T_c$ for $n_f=0$ (left) and $n_f=2$ (right) with various values of $C$. The parameter $\Delta$ is defined as the ratio of the coefficients of the $g^6\ln\,g$ and $g^6$ terms in the expansion (\ref{qcdsusca}). The lattice data is from \cite{gup2,gup3}, and the values $T_c/\Lambda_{{\overline{\rm MS}}}|_{n_f=0}=1.15$ and $T_c/\Lambda_{{\overline{\rm MS}}}|_{n_f=2}=0.49$ \cite{gup1} are used.}
\end{figure}
\begin{figure}[t]

\centerline{\epsfxsize=7.3cm \epsfbox{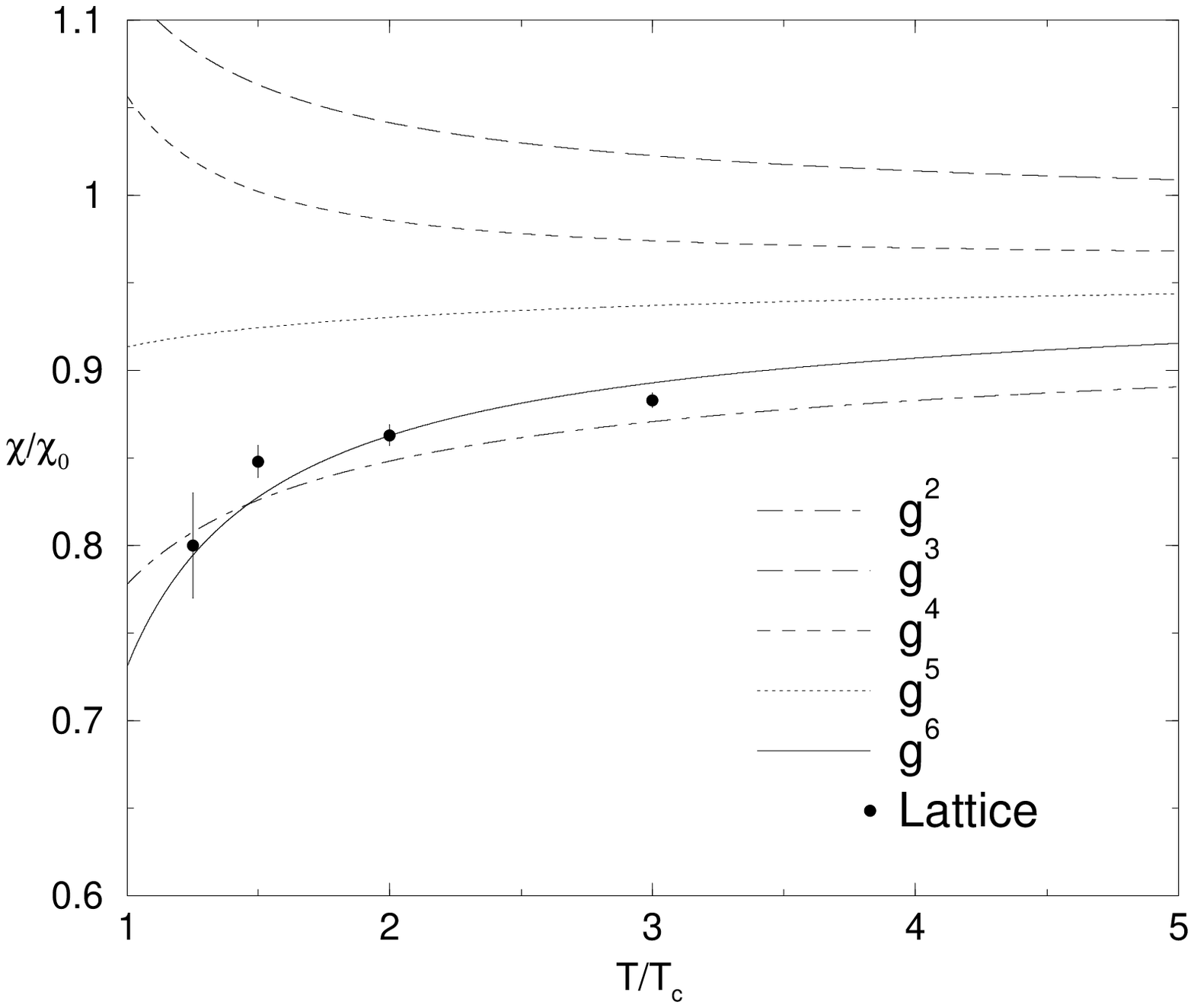}\;\;\;\;\;\;\epsfxsize=7.3cm \epsfbox{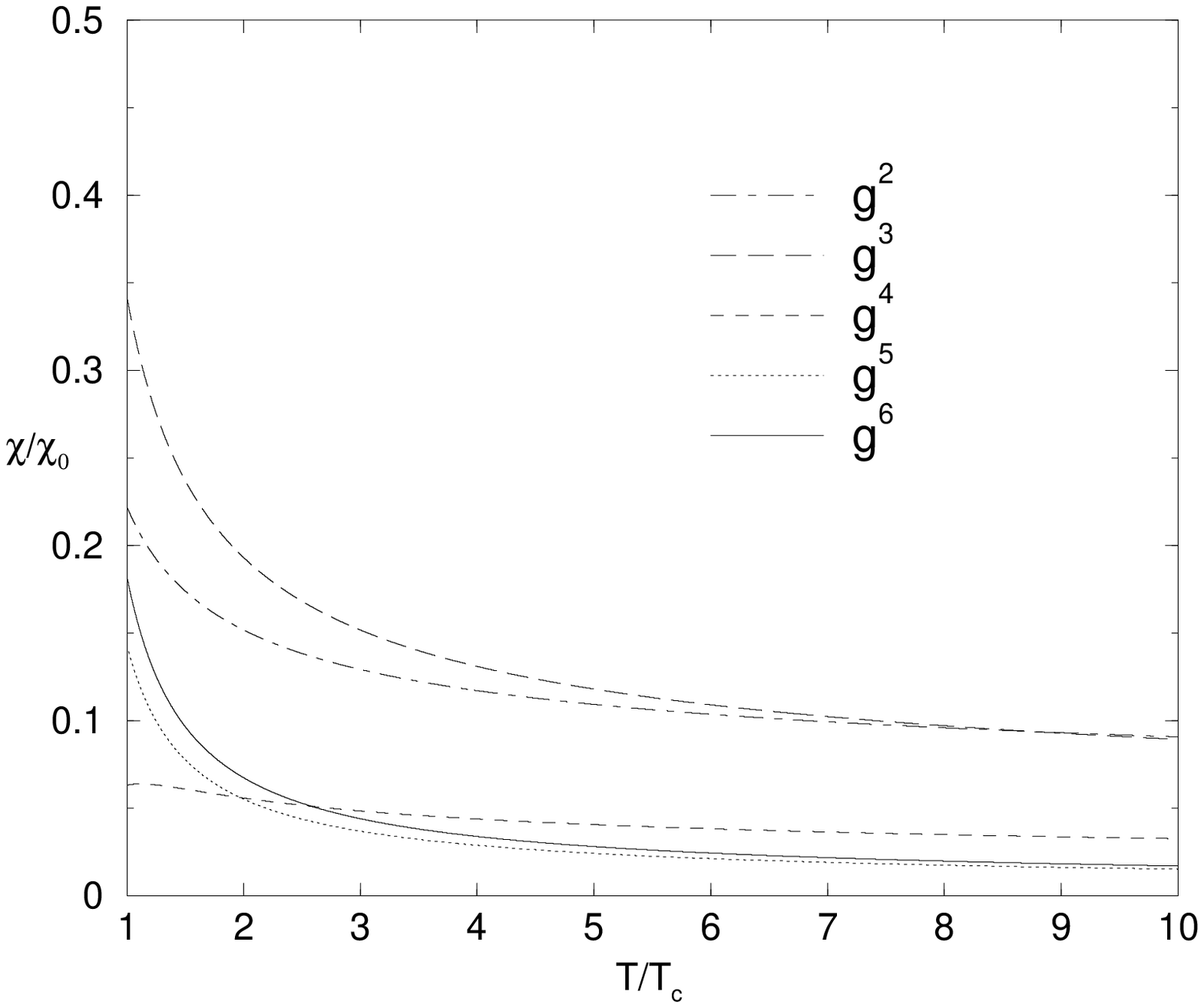}}

\caption[a]{$\chi/\chi_0|_{n_f=2}$ at different perturbative orders for $C(2)=-45$ or $\Delta\approx-0.18$ (left) and the absolute values of the individual terms of the series (right). The lattice results are from \cite{gup3}.}
\end{figure}
\ba
&+& \fr{1}{12\sqrt{1+n_f/6}}\bigg\{\!4\(6+n_f\)\(33-2\,n_f\)\ln\fr{e^{\gamma}\rho}{4\pi T}
-669+30\,n_f+4\,n_f^2 -36\pi^2 \nn
&+& 4\(99-24\,n_f-4\,n_f^2\)\ln\,2+\fr{28}{9}\(6+n_f\)^2\zeta(3)\bigg\}
\bigg(\!\fr{g^2}{4\pi^2}\!\!\!\bigg)^{\!\!5/2} \nn
&+& 2\bigg\{2\(33-2n_f\)\ln\fr{e^{\gamma}\rho}{4\pi T}+\fr{59}{9}+n_f-8\,n_f\ln\,2+\fr{7}{3}\(6+n_f\)
\zeta(3)\bigg\} \bigg(\!\fr{g^2}{4\pi^2}\!\!\!\bigg)^{\!\!3} \ln\,\fr{g^2}{4\pi^2} \nn &-&
\fr{33-2\,n_f}{108}\bigg\{6\(33-2n_f\)\ln\fr{e^{\gamma}\rho}{4\pi T} -
432\,\ln\Big[1+\fr{n_f}{6}\Big]+88+26\,n_f + 16\(17+2\,n_f\)\ln\,2\nn &-& 432\gamma -
432\fr{\zeta'(-1)}{\zeta(-1)}-\fr{2889}{33-2\,n_f}\bigg\}
\bigg(\!\fr{g^2}{4\pi^2}\!\!\!\bigg)^{\!\!3}\!\ln\fr{e^{\gamma}\rho}{4\pi T} +C(n_f)\bigg(\!\fr{g^2}{4\pi^2}\!\!\!\bigg)^{\!\!3} + \mathcal{O}(g^7),
\ea
where $\chi_0 = T^2$ is the free theory result. The two-loop running of $g$
\ba
g^2\!\(\Lambda\) &=&
g^2\!\(\rho\)\bigg[1+\fr{1}{6}\Big(11\,C_A-4\,T_F\Big)\ln\fr{\rho}{\Lambda}\fr{g^2\!\(\rho\)}{4\pi^2} \nn
&+&\fr{1}{12}\!\(17\,C_A^2-10\,C_AT_F-6 \fr{d_FC_F^2}{d_A} +
\fr{1}{3}\(11\,C_A-4\,T_F\)^2\ln\fr{\rho}{\Lambda}\)
\ln\fr{\rho}{\Lambda}\bigg(\!\fr{g^2}{4\pi^2}\!\!\!\bigg)^{\!\!2}\,\bigg] \label{rengroupa}
\ea
has been used to determine the form of the second but last term, which cancels the scale-dependence of the lower-order contributions. The symbol $\rho$ is used in (\ref{qcdsusca}) in place of the scale of dimensional regularization to emphasize the fact that its value is arbitrary, though one expects it to be of order $2\pi T$.

The last term of (\ref{qcdsusca}), proportional to $C(n_f)$, represents the yet undetermined $\mathcal{O}(g^6)$ contribution to the diagonal susceptibility and can only be obtained through a massive computation involving the evaluation of all four-loop diagrams of full QCD contributing to the pressure. It is, however, worth noting that unlike in the case of the pressure \cite{klry} no lattice simulations will be required in this process due to the fact that at order $g^6$ the contribution of $p_\rmi{G}$ to $p_\rmi{QCD}$ is $\mu$-independent. It is therefore obvious that $C(n_f)$ has no direct relation to the coefficient $\delta$ defined in \cite{klry} to represent the $\mathcal{O}(g^6)$ part of $p_\rmi{QCD}$.

In Figs. 2 and 3 the result (\ref{qcdsusca}) is plotted as a
function of $T$ together with lattice data obtained from
\cite{gup2,gup3}. Fig. 2 contains plots of the $n_f = 0$ and $n_f
= 2$ cases for different $C(n_f)$ showing that for reasonable
values of $C$ the perturbative series can be made to approach the
lattice results even at temperatures surprisingly close to $T_c$.
This should, however, by no means be considered an argument
suggesting that the region of applicability of the present results
can be extended down to $T_c$. The success of these perturbative
predictions at $T\leq 3T_c$ is in large part merely a consequence
of having a free parameter available for plotting, and the
behaviour of the result at these temperatures may be completely
distorted by the eventual computation of $C(n_f)$. For reference,
the susceptibility for $n_f=2$ and $C(2)=-45$ is nevertheless
plotted in Fig. 3 to different perturbative orders alongside with
the absolute values of the different terms of the series. For
temperatures higher than a few times $T_c$ the convergence
properties of the series appear to be reasonably good. In the
plots the value of the scale parameter $\rho$ is set to $6.742T$
for $n_f=0$ and to $8.112T$ for $n_f=2$, for which the one-loop
corrections to $g_\rmi{E}^2$ vanish \cite{klrs}.

A few words about the curious concept of susceptibility at zero quark flavours are probably in order. The quantity has been computed
above as the formal $n_f\,\rightarrow\,0$ limit of the general result (\ref{qcdsusca}) and has been plotted in Fig. 2.
The interest in this unphysical limit is due to the existence of recent lattice results from quenched QCD \cite{gup2}
that provide an interesting and powerful test for the validity of the perturbative expansion of the susceptibility. In
particular, being able to compare the perturbation theory results and the lattice data for different numbers of fermion
flavours makes it possible to qualitatively study the $n_f$-dependence of the yet unknown coefficient $C(n_f)$.

\section{Conclusions}
In this paper the diagonal quark number susceptibility of QCD at vanishing chemical potentials has been computed to order $g^6\ln\,g$ in perturbation theory. This is a three-order improvement to the previous result \cite{tt}. Since the next term in the series requires the determination of the perturbative part of the $\mathcal{O}(g^6)$ pressure at finite $T$ and $\mu$, and even the corresponding $\mu=0$ computation seems to be out of reach for current computational techniques, the result obtained here will most likely not be subject to improvement in near future. Nevertheless, a need for further work aiming at the determination of the $\mathcal{O}(g^6)$ terms in the expansions of both the pressure and the susceptibilities is obviously present.

The result obtained here for the diagonal susceptibility resembles the lattice data available but, as demonstrated in the previous sections, may also be considerably modified by the yet undetermined contributions to $\chi$ revealing its lack of predictive power. This could actually have been anticipated due to similar problems encountered in connection with the perturbative expansion of the pressure. The problem can be viewed as a natural consequence of the relatively large value of the QCD coupling constant at the energy scale considered here and of the slow convergence of the perturbative expansion of the pressure in the 3d sector.

While the high temperature, small chemical potential region of the QCD phase diagram is without doubt of considerable special interest, it is also worthwhile to study how the results obtained there can be generalized to other parts of the $(T,\mu)$ -plane. An especially interesting task waiting to be tackled is the building of a bridge between the order $g^4$ perturbative results for the pressure at $T\neq 0,\;\mu=0$ \cite{az} and $T=0,\; \mu \neq 0$ \cite{fmcl}. In the $T\ll\mu$ region the true ground state of the theory may naturally be modified by the appearance of non-perturbative $qq$-condensates \cite{bl}.

\section*{Acknowledgements}
The author wishes to thank Keijo Kajantie for invaluable advice concerning the writing of the paper. Useful comments from Antti Gynther, Mikko Laine and York Schr\"oder are also gratefully acknowledged. The work was supported by the V\"ais\"al\"a Foundation and the Academy of Finland, Contract no. 77744.


\appendix
\renewcommand{\thesection}{Appendix~\Alph{section}}
\renewcommand{\thesubsection}{\Alph{section}.\arabic{subsection}}
\renewcommand{\theequation}{\Alph{section}.\arabic{equation}}

\section{The matching coefficients}
The values of the matching coefficients $\alpha$ defined in chapter 3 read
\ba
\aG &=& \fr{43}{96}-\fr{157}{6144}\pi^2, \\
\aM{1} &=& \fr{43}{32}-\fr{491}{6144}\pi^2, \\
\aM{2} &=& -\fr{2}{3}\bigg(\fr{1}{n_f}\!\sum_f \mubar\bigg)^2, \\
\aE{1} &=& \fr{\pi^2}{45}\fr{1}{n_f}\!\sum_f\bigg\{d_A+d_F\Big(\fr{7}{4} + 30\mubar^2\Big)\!\bigg\},\label{alphae1} \\
\aE{2} &=& -\fr{d_A}{144}\fr{1}{n_f}\!\sum_f\bigg\{C_A + \fr{T_F}{2}\(5+72\mubar^2\)\!\bigg\}, \\
\aE{3} &=& \fr{d_A}{144}\fr{1}{n_f}\!\sum_f\bigg\{C_A^2\bigg[\fr{12}{\e}
+\fr{194}{3}\ln\fr{\bar{\Lambda}}{4\pi T} + \fr{116}{5} + 4\gamma -\fr{38}{3}\fr{\zeta'(-3)}{\zeta(-3)} +
\fr{220}{3}\fr{\zeta'(-1)}{\zeta(-1)}\bigg] \nn
&+& C_A T_F\bigg[ \!12\(1+12\mubar^2\)\fr{1}{\e} +
\Big(\fr{169}{3}+600\mubar^2\Big)\ln\fr{\bar{\Lambda}}{4\pi T} +\fr{1121}{60} -\fr{157}{5}\ln\,2+ 8\gamma \nn
&-& 2\(91-156\gamma-176\,\ln\,2\)\mubar^2
- \fr{1}{3}\fr{\zeta '(-3)}{\zeta(-3)} + \fr{2}{3}\(73+432\mubar^2\)\fr{\zeta'(-1)}{\zeta(-1)}\bigg] \nn
&+& C_F T_F \bigg[\fr{105}{4}-24\,\ln\,2+6\(35+16\,\ln\,2\)\mubar^2\bigg] \nonumber
\ea
\ba
&+& T_F^2 \bigg[\Big(\fr{20}{3}+96\mubar^2\Big)\ln\fr{\bar{\Lambda}}{4\pi T}
+ \fr{1}{3}-\fr{88}{5}\ln\,2 + 4\gamma - 8\(13+16\,\ln\,2-12\gamma\)\mubar^2 \nn
&-& \fr{8}{3}\fr{\zeta'(-3)}{\zeta(-3)} + \fr{16}{3}\fr{\zeta'(-1)}{\zeta(-1)} \bigg]\bigg\}, \label{alphae3} \\
\aE{4} &=& \fr{1}{3}\fr{1}{n_f}\!\sum_f\Big\{C_A+T_F\big[1+12\mubar^2\big]\Big\}, \\
\aE{5} &=& \fr{1}{3}\fr{1}{n_f}\!\sum_{f}
\bigg\{2\,C_A\bigg[\ln\fr{\bar{\Lambda}}{4\pi T} + \fr{\zeta'(-1)}{\zeta(-1)}\bigg]
+ T_F\bigg[2\(1+12\mubar^2\)\ln\fr{\bar{\Lambda}}{4\pi T} + 1 - 2\,\ln\,2 \nn
&-& 12\(1-4\,\ln\,2-2\gamma\)\mubar^2+2\fr{\zeta'(-1)}{\zeta(-1)}\bigg]\bigg\}, \\
\aE{6} &=& \fr{1}{9}\fr{1}{n_f}\!\sum_f\bigg\{C_A^2\bigg[22\,\ln\fr{e^{\gamma}\bar{\Lambda}}{4\pi T}+5\bigg]
+C_AT_F\bigg[2\(7+132\mubar^2\)\ln\fr{e^{\gamma}\bar{\Lambda}}{4\pi T}+9-16\,\ln\,2+4\(33+14\,\zeta(3)\)\mubar^2\bigg] \nn
&-& 18\,C_F T_F\Big[1+12\mubar^2\Big] - 4\,T_F^2\(1+12\mubar^2\)\bigg[2\,\ln\fr{e^{\gamma}\bar{\Lambda}}{4\pi T} - 1 +4\,\ln\,2 - 14\,\zeta(3)\mubar^2 \bigg]\bigg\}, \\
\aE{7} &=& \fr{1}{3}\fr{1}{n_f}\!\sum_{f}\bigg\{C_A\bigg[22\,\ln\fr{e^{\gamma}\bar{\Lambda}}{4\pi T}
+1\bigg] - 4\,T_F\bigg[2\,\ln\fr{e^{\gamma}\bar{\Lambda}}{4\pi T}+4\,\ln\,2-14\,\zeta(3)\mubar^2\bigg]\bigg\}.
\ea



\end{document}